\documentclass[twocolumn]{article}

\usepackage{cite}
\usepackage{amsmath,amssymb,amsfonts}
\usepackage{algorithmic}
\usepackage{graphicx}
\usepackage{textcomp}
\usepackage{xcolor}
\usepackage{booktabs}
\usepackage[affil-it]{authblk} 
\usepackage[margin=0.75in]{geometry} 
\usepackage{tabularx}
\usepackage{titlesec}

\titlespacing*{\section}{0pt}{2.5ex plus 1ex minus .2ex}{1.5ex plus .2ex}
\titlespacing*{\subsection}{0pt}{2ex plus 1ex minus .2ex}{1ex plus .2ex}
\titlespacing*{\subsubsection}{0pt}{1.5ex plus 1ex minus .2ex}{0.5ex plus .2ex}

\title{\textbf{Voice-guided Orchestrated Intelligence for Clinical Evaluation (VOICE): A Voice AI Agent System for Prehospital Stroke Assessment}}

\author[1]{Julian N. Acosta, MD}
\author[2]{Scott J. Adams, MD}
\author[3]{Julius M. Kernbach, MD}
\author[1]{Romain Hardy, MS}
\author[1]{Sung Eun Kim, MD}
\author[1]{Luyang Luo, PhD}
\author[1]{Xiaoman Zhang, PhD}
\author[1]{Shreya Johri, BS}
\author[1]{Mohammed Baharoon, BS}
\author[1,*]{Pranav Rajpurkar, PhD}

\affil[1]{Department of Biomedical Informatics, Harvard Medical School, Boston, MA, USA}
\affil[2]{Department of Medical Imaging, University of Saskatchewan, Saskatoon, SK, Canada}
\affil[3]{Department of Neuroradiology, University Hospital Heidelberg (UKHD), Heidelberg, Germany}
\affil[*]{\textit{ julian\_acosta@hms.harvard.edu}}

\date{}

\begin{document}

\maketitle

\begin{abstract}
We developed a voice-driven artificial intelligence (AI) system that guides anyone—from paramedics to family members—through expert-level stroke evaluations using natural conversation, while also enabling smartphone video capture of key examination components for documentation and potential expert review. This addresses a critical gap in emergency care: current stroke recognition by first responders is inconsistent and often inaccurate, with sensitivity for stroke detection as low as 58\%, causing life-threatening delays in treatment. Three non-medical volunteers used our AI system to assess ten simulated stroke patients, including cases with likely large vessel occlusion (LVO) strokes and stroke-like conditions, while we measured diagnostic accuracy, completion times, user confidence, and expert physician review of the AI-generated reports. The AI system correctly identified 84\% of individual stroke signs and detected 75\% of likely LVOs, completing evaluations in just over 6 minutes. Users reported high confidence (median 4.5/5) and ease of use (mean 4.67/5). The system successfully identified 86\% of actual strokes but also incorrectly flagged 2 of 3 non-stroke cases as strokes. When an expert physician reviewed the AI reports with videos, they identified the correct diagnosis in 100\% of cases, but felt confident enough to make preliminary treatment decisions in only 40\% of cases due to observed AI errors including incorrect scoring and false information. While the current system's limitations necessitate human oversight, ongoing rapid advancements in speech-to-speech AI models suggest that future versions are poised to enable highly accurate assessments. Achieving human-level voice interaction could transform emergency medical care, putting expert-informed assessment capabilities in everyone's hands.
\end{abstract}

\section{Introduction}
Stroke is a leading cause of death and disability \cite{b1}. Yet prehospital stroke recognition remains highly inconsistent, with dispatcher sensitivity rates as low as 41\% \cite{b2} and EMS provider sensitivity varying between 58\% and 76\% \cite{b3,b4}. This variability prevents many patients from receiving time-sensitive interventions, worsening both mortality and long-term disability \cite{b5,b6}. Despite revolutionary advances in acute ischemic stroke treatment, many patients remain ineligible for these time-sensitive therapies due to fundamental failures in stroke recognition and triage decision-making \cite{b5,b6}. 

Existing efforts to improve prehospital stroke assessment, including manual tools like FAST (Face, Arm, Speech, Time) \cite{b7}, FAST-ED (Field Assessment Stroke Triage for Emergency Destination) \cite{b8}, and CPSS (Cincinnati Prehospital Stroke Scale) \cite{b9}, alongside their digital counterparts \cite{b10,b11,b12,b13}, have not fully resolved this problem. These tools, while valuable, often require expert interpretation to navigate ambiguities and systematically underrepresent certain stroke symptoms like those from the posterior circulation \cite{b14,b15}. Current digital solutions largely digitize these static, checklist-driven approaches, maintaining rigid workflows that demand prior knowledge and offer little to no dynamic guidance for complex or uncertain presentations \cite{b10,b11,b12,b13}. Crucially, they typically require users to download, learn, and interact with an app-based interface, a significant departure from natural human communication modalities like voice and video calls, thereby limiting their intuitive use and widespread adoption, especially in high-stress emergency scenarios.

Generative artificial intelligence (AI) and real-time speech processing now make it possible to imagine a new kind of prehospital stroke assessment tool: voice-based conversational agents that could speak, adapt, and reason like experienced clinicians. Unlike rigid digital checklists or unnatural text-based apps \cite{b14}, these systems have the potential to engage users in natural, real-time dialogue---asking follow-up questions, interpreting vague or imprecise responses, and guiding the assessment dynamically. By enabling more intuitive interactions, they could support users with varied levels of medical training, reduce cognitive burden in high-stress scenarios, and standardize stroke triage across settings. These systems could also generate auditable reports of these interactions, helping bridge communication gaps between prehospital and in-hospital care. 

We present the first evaluation of Voice-guided Orchestrated Intelligence for Clinical Evaluation (VOICE), a voice AI agent system designed initially for prehospital stroke assessment, offering early evidence that real-time voice-guided AI may pave the way for more intelligent, accessible, and reliable prehospital triage.

\section{METHODS}

\subsection{Study Design}

This investigation was conducted as a simulation-based feasibility study. All layperson users and patient actors (students or scholars) provided informed consent prior to their participation.

\begin{figure*}[!t]
\centerline{\includegraphics[width=0.75\textwidth]{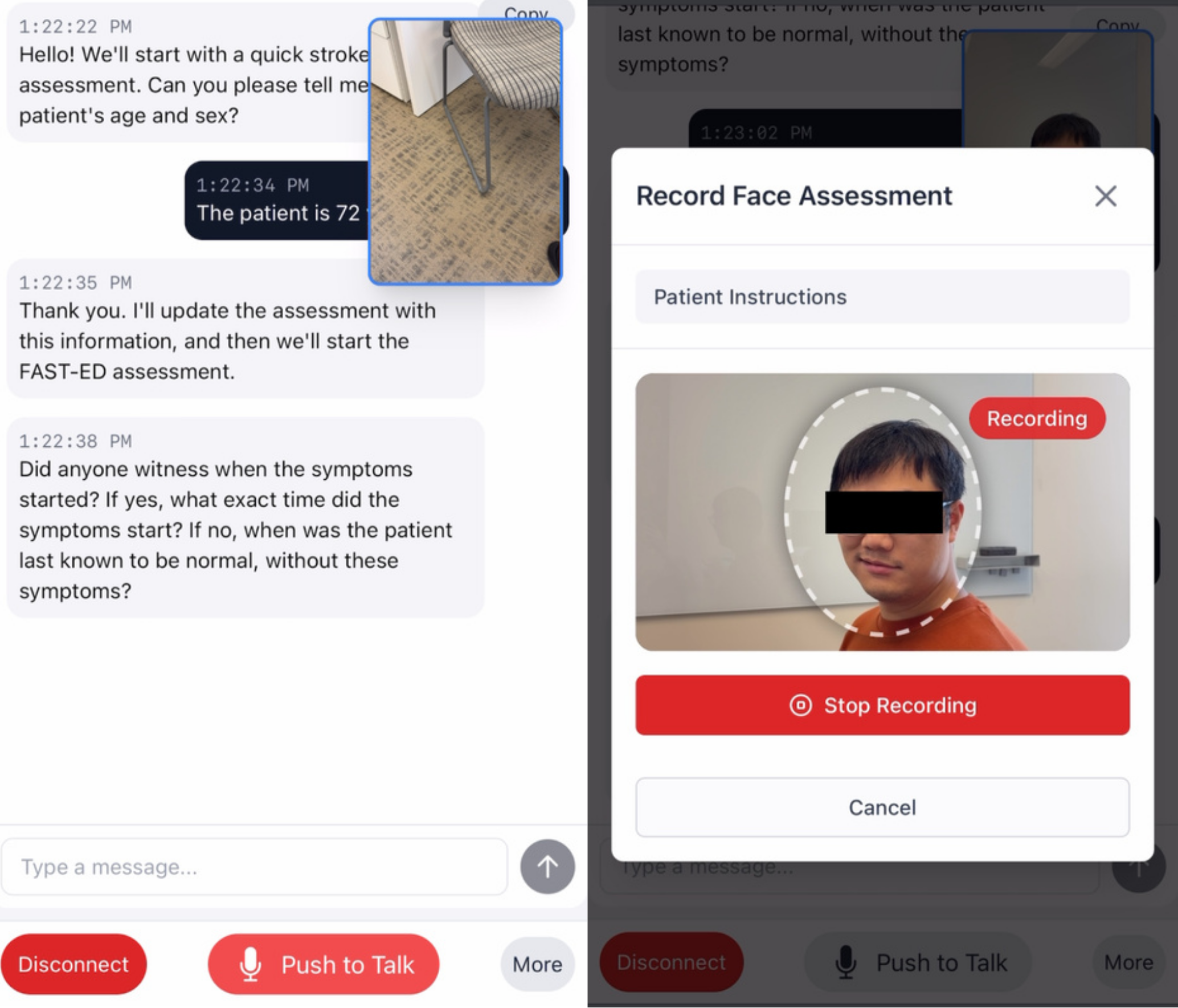}}
\caption{Mobile webapp during an assessment. On the left, during the initial portion of the assessment, the user is pressing "Push to Talk" to respond to the AI via voice. On the right, the user is being guided to record a video for the facial palsy assessment.}
\label{fig1}
\end{figure*}

\begin{figure*}[!t]
\centerline{\includegraphics[width=0.85\textwidth]{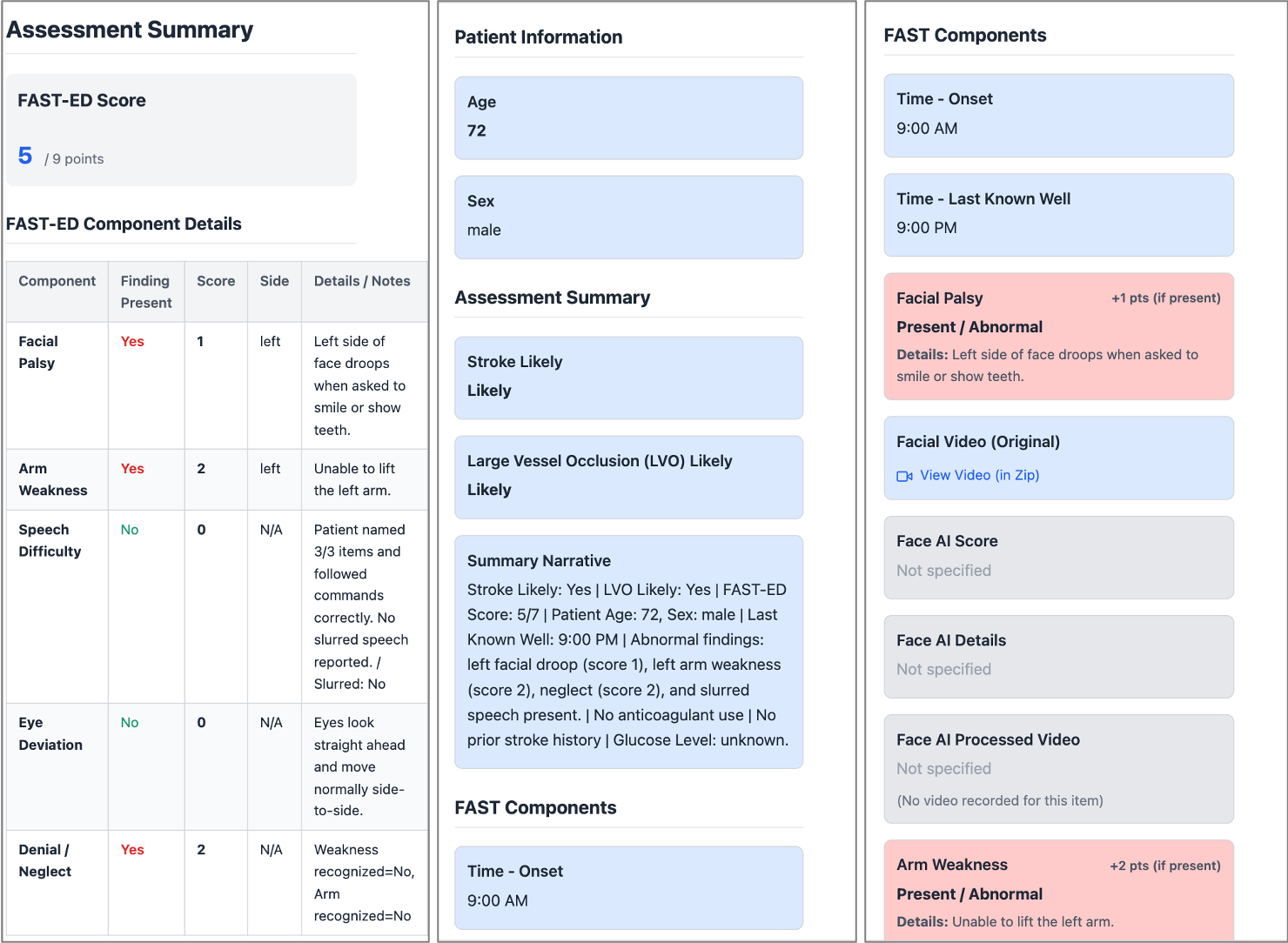}}
\caption{Example report produced by the system (along with the conversation transcription).}
\label{fig2}
\end{figure*}

\begin{table*}[!tb] 
    \renewcommand{\arraystretch}{1.1} 
    \centering 
    \footnotesize
    \caption{Conversation transcript for one simulated patient scenario.} 
    \label{tab:conversation} 
    \begin{tabularx}{\textwidth}{p{1.2cm} X} 
    \toprule 
    \textbf{Speaker} & \textbf{Dialogue} \\ 
    \midrule
\textbf{Assistant} & Hello! We'll start with a quick stroke assessment. Can you please tell me the patient's age and sex? \\
\textbf{User} & The patient is 72 years old and male. \\
\textbf{Assistant} & Thank you. I'll update the assessment with this information, and then we'll start the FAST-ED assessment. \\
\textbf{Assistant} & Did anyone witness when the symptoms started? If yes, what exact time did the symptoms start? If no, when was the patient last known to be normal, without these symptoms? \\
\textbf{User} & The patient was normal around 9 p.m. yesterday and then this morning during a video call I noticed that his face was drooping. \\
\textbf{Assistant} & To be precise, can you give me a specific clock time for when the symptoms were noticed during the video call this morning? If not, I'll use 9:00 PM yesterday as the last known well time. \\
\textbf{User} & around 9 a.m. \\
\textbf{Assistant} & Please record a video of the patient's facial movements. Ask the patient to smile or show their teeth, and let me know if both sides of the face move equally or if one side droops more than the other. \\
\textbf{User} & The left side of his face droops. \\
\textbf{Assistant} & Please ask the patient to raise and extend both arms with palms up and hold for exactly 10 seconds. Watch closely and let me know if either arm drifts down before the time is up, falls rapidly, or cannot move at all. \\
\textbf{User} & The patient is able to lift up his right arm but is unable to lift up his left arm. \\
\textbf{Assistant} & We're assessing speech now. Is the patient's speech slurred or unclear? Please ask them to repeat ``Today is a sunny day'' and let me know if their speech is slurred or not. \\
\textbf{User} & His speech is now slurred. \\
\textbf{Assistant} & Now, let's test for aphasia. Please ask the patient to name three common items, like a pen, a watch, and a button. Let me know how many items they can name correctly. \\
\textbf{User} & The patient is able to name three common items. \\
\textbf{Assistant} & Now, please ask the patient to perform a simple command, like ``show me two fingers.'' Let me know if they can perform the command correctly. \\
\textbf{User} & Yes, you can perform the command. \\
\textbf{Assistant} & Look carefully at the person's eyes. Are they stuck looking all the way to the left, or all the way to the right? Or can they look straight ahead and move their eyes side-to-side? Ask the patient to follow your finger while moving it to the left and to the right to evaluate this. \\
\textbf{User} & The patient is able to look straight ahead and can follow my finger just fine. \\
\textbf{Assistant} & Okay, we need to check the patient's awareness of the weakness. First, please show the patient their weak arm and ask them directly: ``Is this arm weak?'' Note if they recognize the weakness or do not recognize it. Second, keeping focus on the weak arm, ask the patient directly: ``Whose arm is this?'' Note if they recognize it as their own arm or do not recognize it. Please tell me the results for both questions: Did they recognize the weakness? Did they recognize the arm? \\
\textbf{User} & The patient cannot recognize the weakness and do not recognize the arm as their own. \\
\textbf{Assistant} & Is the patient taking any of these anticoagulant medications: Coumadin/Warfarin, Pradaxa/Dabigatran, Eliquis/Apixaban, Xarelto/Rivaroxaban, Savaysa/Edoxaban, or Heparin/Enoxaparin? \\
\textbf{User} & He's not taking any of these medications. \\
\textbf{Assistant} & I need a couple more pieces of information. Does the patient have a history of a previous stroke? If yes, can you provide any details? \\
\textbf{User} & Uh, not that he's aware of. \\
\textbf{Assistant} & Do you have any way to check the patient's blood glucose (blood sugar) level? If so, please provide the level. If not, just let me know. \\
\textbf{User} & No, we're unable to measure it, we don't know. \\
\textbf{Assistant} & I'm processing the results now, this may take a minute. \\
\textbf{Assistant} & I need to clarify one thing: Is the patient male or female? I have a discrepancy in the records. \\
\textbf{User} & The patient is male. \\
\textbf{Assistant} &  The assessment indicates that a stroke is likely, and there's a high likelihood of a large vessel occlusion (LVO). The FAST-ED score is 5 out of 7, with abnormal findings including left facial droop, left arm weakness, neglect, and slurred speech. If you have not already, you need to call 911 as soon as possible and indicate a possible stroke. \\
\textbf{Assistant} & The assessment summary has been provided. Is there anything else I can help you with before disconnecting? \\
\textbf{User} & Huh, no. \\
    \bottomrule 
    \end{tabularx} 
\end{table*}

\subsection{AI System Architecture: Core Technology and Agents
}

The system was developed as a web application using Next.js for the front-end, based on a modified version of the code provided by OpenAI at https://github.com/openai/openai-realtime-agents. Modifications included integration with an ad-hoc relay server to utilize Azure OpenAI services and the use of WebSockets for real-time communication (as Azure OpenAI does not currently support WebRTC). The web application was hosted on secure Azure environments. The core of the system utilized the OpenAI Realtime API with the \textit{gpt-4o-realtime-preview} model (version 2024-12-17). This model directly consumes audio and text, maintains session context, and produces streaming audio and text outputs via the WebSocket connection. While the model processes audio directly, the OpenAI Realtime API also performs parallel transcription using \textit{whisper-1} for displaying user messages on the interface. Users interacted with the system primarily via voice using a "push-to-talk" button on the web interface; visual feedback indicated when the microphone was active (Figure~\ref{fig1}). At the end of interactions, the system generates a comprehensive report which includes component-wise and total FAST-ED (a medical scale designed for prehospital stroke assessment \cite{b8}) scores, likelihood of stroke, likelihood of large vessel occlusion (LVO) stroke, all collected ancillary data, links to the recorded videos, and the full conversation transcript (Figure~\ref{fig2}, Table~\ref{tab:conversation}).

The system employed a multi-agent architecture\cite{b16}. A Main Orchestration Agent managed the overall assessment flow, initiating the interaction, collecting initial demographic data, handing off to specialized component agents, collecting final ancillary data, and managing session termination. Specialized Component Assessment Agents, each with specific instructions and tools, handled distinct parts of the assessment. These included agents for determining Symptom Onset and Last Known Well Time, Facial Palsy assessment, Arm Weakness assessment, Speech Abnormality assessment, Eye Deviation assessment, Denial/Neglect assessment, and Anticoagulant Use inquiry. Each component agent guided the user through its specific task, processed responses, and updated a shared assessment state. A predefined assessment schema structured the storage of all collected data.

A Final Summary Agent was invoked towards the end of the assessment. Its primary role was to initiate a comprehensive analysis by a reasoning LLM (\textit{o3-mini}, version 2025-01-31). This analysis involved reviewing the entire collected assessment state and conversation transcript to double check the final FAST-ED component and total scores, determine stroke and LVO likelihood, and draft a concise summary narrative. The Final Summary and Analysis Agent then updated these findings into the final assessment state if needed, and/or instructed the Main Agent to clarify any needed information. The flow between agents was predefined, with each agent's instructions guiding the transfer to the next appropriate agent. The assessment state and conversation context were maintained throughout the session, with the active agent's specific instructions loaded dynamically.

The AI agents utilized several key system-level functions or ``tools'' defined in their operational instructions. An Update Assessment State function was called by agents to record specific findings into a centralized, dynamic assessment state. Video Recording Initiation functions (for facial and arm assessments) were called by the respective component agents before instructing the user on the physical exam, opening a modal for video recording with on-screen controls; successful recording returned a video identifier. A Transfer Agent Control function was called by an agent upon task completion to hand off control to the next predefined agent. A specialized function was also  called only by the Final Summary Agent to invoke \textit{o3-mini} for double checking scores, final score calculation, and summary drafting. A Disconnect Session function was called by the Main Orchestration Agent at the end of the interaction.

\subsection{Patient Actor Scenarios and Ground Truth Establishment}
Ten unique standardized patient actor scripts were developed by a neurologist. These scripts covered a range of clinical presentations, including likely ischemic stroke cases, likely LVOs, and likely non-ischemic stroke conditions. Nonischemic stroke scenarios included presentations such as confusion due to a serious nonneurological illness (infection), a migraine aura, and a thunderclap headache without focal neurological symptoms. Each script provided detailed information about the patient's history, symptoms, and specific signs to be enacted for each FAST-ED component. The ground truth for each scenario, including FAST-ED component scores, total FAST-ED score, overall stroke/non-stroke status, and LVO status, was defined based on the neurologist-developed case details actors were instructed to portray.

Three layperson users (students or scholars) participated in the study. Prior to starting the AI-guided assessments, users were provided with "Caller Scenario Cards" for each case. These cards contained contextual information about the simulated situation and what the "caller" (the user) had purportedly noticed before initiating communication with the AI system. The patient actor scripts contained more detailed information regarding the patient's history, symptoms, and signs, which the actors used for their performance. Users were shown the web application and its interface before initiating the cases and were aware they were interacting with an AI system. They were instructed to follow the AI's voice prompts to assess the patient actor and then verbally report their direct observations back to the AI.

\subsection{Outcome Measures}
The study evaluated the system across several domains: diagnostic accuracy of the AI-guided assessment, operational efficiency, physician reception of system outputs, and user experience.

\noindent \textbf{AI-Guided Assessment Accuracy:} This included the Accuracy of FAST-ED Component Scoring, determined by the concordance of each AI-guided FAST-ED component score (facial palsy, arm weakness, speech/aphasia, eye deviation, denial/neglect) and the total FAST-ED score with the established ground truth for each scenario. Accuracy of Stroke and LVO Identification was assessed by comparing the AI-system's determinations against the Ground Truth status. Accuracy of Ancillary Data Collection evaluated the correctness of AI-elicited information (patient age, gender, Last Known Well time, symptom onset time, anticoagulant use, prior stroke history, glucose level) against script-defined details. Finally, a qualitative analysis of AI assessment discrepancies was performed by reviewing reports, transcripts, and videos to identify and categorize issues such as component mis-scoring, AI hallucinations (information reported by AI not stated by user/actor or in scripts), and reasons leading to overcalling stroke in mimics, and missed stroke diagnoses.

\noindent  \textbf{Operational Efficiency:} Assessment Completion Time was measured as the mean time from the start of the user-AI interaction to the AI's delivery of the final assessment summary.

\noindent \textbf{Physician Evaluation:} A physician experienced in evaluating and treating stroke patients, blinded to the ground truth, reviewed all 10 AI-generated reports (containing the AI-assigned FAST-ED components, transcript, and videos for the face and arm weakness components). The physician adjudicated the FAST-ED score from the report, which was then compared to the AI-guided scores and ground truth. The physician also assessed report Completeness (whether reports contained all necessary FAST-ED elements and key ancillary data). Additionally, we measured the physician's confidence (5-point Likert scale: 1=Not at all confident to 5=Extremely confident) in using the complete AI report for preliminary clinical decisions (such as referring the patient to a comprehensive stroke center directly), with written justification. Qualitative feedback collected the physician's insights on report quality, video utility, and suggestions.

\noindent  \textbf{User Experience:} Per-Case Assessment Confidence was rated by users after each scenario on a 5-point Likert scale (1=Not at all confident to 5=Extremely confident) regarding the correctness of their AI-guided assessment. A Post-Study Usability Survey collected overall Ease of Voice Interaction (5-point Likert scale: 1=Very Difficult to 5=Very Easy) and Qualitative Feedback through open-ended questions on likes, challenges (including speech-to-text issues), and suggestions for system improvement.

\subsection{Statistical Analysis}
Quantitative data (concordance rates, accuracy percentages, assessment times, Likert scores) were summarized using descriptive statistics (frequencies, percentages, means, SD, median, IQR). Qualitative data (user/physician feedback, AI discrepancy analysis) underwent thematic analysis.

\section{RESULTS}
\subsection{Stroke Identification and FAST-ED Score Accuracy}
Regarding FAST-ED component scores, accuracy compared to ground truth across the 10 scenarios was as follows: Facial Palsy, 10 of 10 (100\%) correct; Arm Weakness, 8 of 10 (80\%) correct; Aphasia, 8 of 10 (80\%) correct; Eye Deviation, 8 of 10 (80\%) correct; and Denial/Neglect, 8 of 10 (80\%) correct. This resulted in 42 out of 50 (84\%) individual FAST-ED component findings being accurately captured by the AI-guided user reports.

The total AI-guided FAST-ED score matched the ground truth score in 5 out of 10 cases (50\%). For the remaining 5 cases, in 3 cases (30\%), the AI-derived total score differed by $\pm$1 point from the ground truth, and in 2 cases (20\%), the discrepancy was $\pm$2 points.

For overall stroke classification, the voice AI agent system correctly identified 6 of the 7 true stroke cases and incorrectly classified 2 of the 3 non-stroke mimic scenarios as strokes (sensitivity 86\%; specificity 33\%; accuracy 70\%). The AI system correctly identified 3 out of 4 LVO cases (sensitivity 75\%; specificity 100\% based on non-LVO stroke cases; overall accuracy 86\% across all cases when considering LVO vs. non-LVO/mimic).

The AI system accurately captured patient gender, prior stroke history, and glucose values in all 10 cases (100\%). Patient age, last known well time, and symptom onset time were correctly recorded in 9 out of 10 cases (90\%); in one instance, the system recorded symptom onset time rather than last known well time when both were provided by the user. Anticoagulant history was correctly reported in 8 out of 10 cases (80\%); one of these correctly identified anticoagulant use but missed the last dose taken, and in another case, an anticoagulant was hallucinated by the AI based on the user mentioning an unrelated medication (diabetes medication).

\subsection{Assessment Time}
The mean time to complete the AI-guided assessments, from the initiation of the interaction to the AI delivering its final summary, was 6 minutes and 15 seconds (Standard Deviation [SD] 42 seconds) across all 10 scenarios.

\subsection{Qualitative Analysis of AI Assessment Discrepancies and Failures}
Discrepancies included one missed true stroke case (a posterior circulation stroke presenting with dizziness and diplopia), likely reflecting the FAST-ED scale's design primarily for anterior circulation LVO strokes. In one LVO case, the AI agent failed to prompt for the denial/neglect component, leading to an incomplete FAST-ED assessment and an incorrectly lower total score. In another instance, the user’s description of the patient’s slow speech was misinterpreted by the user as dysarthria/aphasia. There was also a case where the AI incorrectly assigned arm weakness and subsequently denial/neglect, despite the user explicitly stating that no arm weakness was present.

\subsection{Physician Evaluation}
The reviewing experienced physician evaluating the AI-generated reports (including text and videos), achieved perfect concordance with the ground truth FAST-ED scores in 9 out of 10. In one case, the physician's score was 1 point lower than ground truth due to an AI system incompleteness in the assessment (missed denial/neglect component). The evaluator reached the correct diagnosis of stroke and/or LVO in all cases.

When asked to rate confidence in using the complete AI-generated report (assessment scores, transcript, videos) for preliminary clinical decision-making (1=Not at all confident to 5=Extremely confident), the physician assigned a score of $\geq$3 in only 4 out of 10 cases. Reasons provided for lower confidence ratings (scores $<$3) in the remaining 6 cases included: incorrect AI assessment of aphasia leading to a miscalculated FAST-ED score and potentially erroneous clinical judgment (e.g., interpreting slurred speech as aphasia); AI errors in recording medication (e.g., Xarelto) which would impact decisions on lytic therapy; missing assessment components (denial/neglect); AI misinterpretation of arm weakness or unintelligible transcription of user input leading to incorrect scoring; and concerns that the AI's binary ``stroke likely/unlikely'' output did not adequately capture differential diagnoses (e.g., for a thunderclap headache).

\subsection{User Feedback}
After completing each scenario, users rated their confidence in the correctness of their AI-guided assessment. The median confidence score across all 10 user-case assessments was 4.5 on a 5-point Likert scale (Interquartile Range [IQR] 4--5).

Upon completion of all scenarios, users responded to a survey. For the question ``How easy or difficult was it to interact with the AI system using your voice?'' (1=Very Difficult to 5=Very Easy), two users rated it as 5 (Very Easy) and one user rated it as 4 (mean score 4.67). Qualitative feedback indicated users appreciated the system's clear instructions and rapid, specific questioning. However, challenges included occasional AI hallucinations or inconsistencies, imperfect speech-to-text leading to misunderstandings (e.g., misinterpreting numbers), a perception of the AI being overly rigid in its conversational flow, and its speaking pace sometimes being too slow. Key suggestions for improvement centered on enhancing speech-to-text accuracy, offering more conversational flexibility (such as an always-listening mode), increasing the AI's response speed, providing user reassurance, and enabling mechanisms for correcting input errors.

\section{DISCUSSION}

This study demonstrates the significant potential of a novel, voice-driven AI agent system with integrated video capture to guide users through structured prehospital stroke assessments. The intuitive voice interface, mirroring natural human communication, offers a compelling advantage in lowering barriers to urgent medical evaluations, particularly for individuals unfamiliar with medical jargon or complex apps. Our system showed promise especially in identifying LVO stroke cases and received positive feedback on usability. However, observed limitations—including inaccuracies in FAST-ED component scoring, a tendency to overcall stroke in non-stroke cases, and occasional AI hallucinations—underscore that current speech-to-speech AI models, while enabling fluid conversation, still require significant improvements for robust clinical deployment. These challenges highlight areas where rapid advancements in AI are poised to make a substantial impact.

Existing digital stroke assessment tools---such as mobile apps that digitize the FAST-ED or CPSS scales \cite{b10,b11,b12,b13}---function as static digital checklists, guiding users through a fixed sequence of yes/no questions without adapting to the user's responses or level of uncertainty. These apps typically require manual data entry, lack real-time feedback, and provide minimal support for nuanced clinical reasoning or clarification of ambiguous input. Moreover, they generally do not capture audio or video, limiting their utility for documentation, auditability, or retrospective review. In contrast, the AI agent system introduced in this study engages users through natural spoken dialogue, adapts its questioning based on responses, and integrates video capture to document observable signs like facial droop or arm weakness.

Current performance issues highlighted in our study are largely attributable to the current state of end-to-end speech-to-speech AI. While our system achieves a natural conversational flow by using a model that directly processes and generates audio, this sacrifices the deeper reasoning capabilities seen in more mature text-based Large Language Models (LLMs)\cite{b17}. The alternative paradigm of using speech-to-text (STT), then LLM processing, then text-to-speech (TTS) the LLM output~\cite{b18}, while potentially leveraging more powerful LLMs, often disrupts conversational fluidity and loses the rich nuances (tone, hesitation, emphasis) present in human speech—nuances that can be vital for clinical interpretation and for supporting users with lower literacy levels. Ironically, while designing for more natural, nuanced interaction aims to broaden accessibility, it also introduces complexities that current speech models do not yet fully master. We anticipate that as audio-to-audio (or true speech-to-speech) models rapidly evolve to match the contextual understanding and reasoning of today's leading text-based LLMs (like \textit{o3}, \textit{Gemini 2.5 Pro}, or \textit{Claude 4}), systems like VOICE will achieve far greater accuracy and smoother, more fault-tolerant interactions.

A key strength of our system is its modular architecture and multimodal data capture. While the current implementation is designed to guide human users through structured assessments, the integrated video and audio streams collected during these interactions could be leveraged to train AI models capable of identifying physical signs such as facial asymmetry, slurred speech, arm drift, or gaze deviation. These features are critical in stroke assessment but are currently reliant on human observation and reporting. Building on prior work ~\cite{b19,b20}, we are currently developing automated assessment modules for several of these components, including facial palsy, arm weakness, and gaze deviation based on models to detect facial landmarks and pose, which may eventually enable real-time AI-supported scoring. Future versions of the system may incorporate computer vision models to analyze video in real time\cite{b21}. Achieving this level of autonomy will require substantial advances in multimodal model development, including improvements in contextual reasoning across modalities and robustness to real-world variation in video and audio quality. Nevertheless, the current system lays the necessary groundwork by capturing structured, labeled, and context-rich multimodal data that can support iterative model training and validation over time. Future iterations could also potentially leverage advances in augmented reality\cite{b22} to provide a more natural, hands-free assessment perspective aligned with how clinicians evaluate patients. Furthermore, to maximize accessibility and leverage user familiarity, this voice AI technology could ultimately be embedded within widely adopted communication platforms, such as standard video calling services (akin to FaceTime), obviating the need for dedicated application downloads and user training.

Beyond acute prehospital assessment, the VOICE system also holds considerable promise as an accessible and scalable training tool. Laypersons, family members, and even EMS personnel could use the system in simulated scenarios to practice and enhance their stroke recognition skills. The guided, interactive nature of the AI, coupled with immediate feedback and the ability to review recorded interactions, offers a dynamic learning environment that surpasses static training materials or infrequent practical sessions. This could contribute to a more broadly educated populace and a higher baseline of stroke awareness and assessment capability within communities and professional services.

There are a few limitations to this study. This feasibility study was conducted entirely in a simulated environment using standardized patient actor scenarios. While this controlled setting allows for consistent evaluation and reproducibility, it does not capture the full complexity of real-world prehospital care, including background noise, competing patient needs, and time pressure. FAST-ED scores were determined by a physician, but the absence of real patients means that certain nuances of stroke presentation may not have been fully replicated. The participants assessing the cases had variable non-clinical backgrounds, and none of the participants was a trained paramedic or EMS professional, which may limit the generalizability of findings to users with other backgrounds. Additionally, the sample size of callers and scenarios was limited, and no formal power calculations were performed given the exploratory nature of the study. 

In conclusion, this study establishes the feasibility and highlights the transformative potential of a generative AI voice agent with integrated video for prehospital stroke assessment. By leveraging natural spoken interaction, our system offers a more intuitive, accessible, and potentially more reliable approach than existing static tools, empowering a wider range of users to conduct critical evaluations. While the current system demonstrates promising LVO identification and user acceptance, its full potential will be unlocked by ongoing rapid advancements in audio-to-audio AI models, which are expected to significantly improve reasoning, accuracy, and conversational nuance. The modular design, allowing for future integration of automated video/audio analysis and deployment on familiar communication platforms, positions such systems to redefine prehospital triage, training, and ultimately, emergency medical response by delivering scalable, intelligent, and human-like clinical guidance when and where it is needed most.

\section*{Acknowledgments}

We gratefully acknowledge Sraavya Sambara, Siddhant Dogra, Joao Brainer Clares de Andrade, Matias Alet, Santiago Clocchiatti-Tuozzo, and Cyprien Rivier for their valuable help and feedback on this work. We also thank the Microsoft Accelerating Foundation Models Research program and Harvard Medical School’s Dean’s Innovation Award for facilitating access to compute resources.

\end{document}